\documentclass{l4dc2025}

\usepackage{times}
\usepackage{array}
\usepackage{booktabs}
\usepackage{comment}
\usepackage{caption}
\usepackage{adjustbox}
\usepackage{mathtools}
\usepackage{multicol}
\usepackage{multirow}
\usepackage{hyperref}
\usepackage{xr}

\title[Interaction-Aware Parameter Privacy-Preserving Data Sharing in Coupled Systems]{Interaction-Aware Parameter Privacy-Preserving Data Sharing in Coupled Systems via Particle Filter Reinforcement Learning}

\author{%
  \Name{Haokun Yu} \Email{yuhaokun@u.nus.edu}\\
  \addr Institute of Operations Research and Analytics, National University of Singapore, Singapore 117602
  \AND
  \Name{Jingyuan Zhou} \Email{jingyuanzhou@u.nus.edu}\\
  \addr Department of Civil and Environmental Engineering, National University of Singapore, Singapore 119077
  \AND
  \Name{Kaidi Yang} \Email{kaidi.yang@nus.edu.sg}\\
  \addr Department of Civil and Environmental Engineering, National University of Singapore, Singapore 119077
}

\begin{document}

\maketitle

\begin{abstract}%
This paper addresses the problem of parameter privacy-preserving data sharing in coupled systems, where a data provider shares data with a data user but wants to protect its sensitive parameters. The shared data affects not only the data user's decision-making but also the data provider's operations through system interactions. To trade off control performance and privacy, we propose an interaction-aware privacy-preserving data sharing approach. 
Our approach generates distorted data by minimizing a combination of (i) mutual information, quantifying privacy leakage of sensitive parameters and (ii) the impact of distorted data on the data provider’s control performance, considering the interactions between stakeholders. The optimization problem is formulated into a Bellman equation and solved by a particle filter reinforcement learning (RL)-based approach. Compared to existing RL-based methods, our formulation significantly reduces history dependency and efficiently handles scenarios with continuous state space.  
Validated in a mixed-autonomy platoon scenario, our method  effectively protects sensitive driving behavior parameters of human-driven vehicles (HDVs) against inference attacks while maintaining negligible impact on fuel efficiency. 
Detailed proofs and experiment setup can be found in \href{https://drive.google.com/file/d/1jpP_dgP1uthVSZXIkWZKRVjlXeYXGLDU/view?usp=sharing}{this supplementary material}.

\end{abstract}


\section{Introduction}

In the era of big data, data sharing between stakeholders has become a cornerstone for improving the operations of various cyber-physical systems such as transportation \citep{zheng2018traffic}  and power systems \citep{larsen2014power}. However, since operational data can contain sensitive information about customers and system operations, data sharing can raise significant privacy concerns. For example, although trajectory data of a ride-hailing company can enhance transportation practice and research, it contains sensitive information about individual mobility patterns (e.g., destinations and routing preferences) and about the company's sensitive operational parameters. Such privacy concerns can deter stakeholders from sharing data, thereby undermining the value of big data. 

Despite the recent development of privacy-preserving data-sharing mechanisms, such as differential privacy (DP) \citep{shokri2014privacy,zhao2014achieving,chen2023optimization},  most research focuses on protecting individual-level data, while overlooking the privacy of sensitive parameters. However, sensitive parameters, often treated as business secrets, serve as one of the key reasons that can prevent stakeholders from data-sharing. For example, the algorithmic parameters of ride-hailing companies, if leaked, can compromise the competitive advantages of the company. In control systems, the leakage of feedback gains~\citep{nekouei2021randomized} and internal states~\citep{nekouei2022model,weng2023optimal} can allow malicious actors to exploit vulnerabilities, causing system failures. These examples underscore the critical need to protect sensitive parameters, as their exposure can result in operational disruptions, competitive disadvantages, and security threats.

The research considering parameter privacy is sparse. 
Popular techniques such as differential privacy and k-anonymity seek to protect individual privacy by hiding them in a herd. Nevertheless, there is no herd to hide when it comes to protecting the parameter privacy. To address this issue, information-theoretic methods have attracted increasing attention, which use metrics like mutual information and entropy to precisely quantify the privacy leakage of parameters by sharing data. For example, \citet{bassi2018lossy} uses stochastic kernels to protect statistical properties involving private parameters, and \citet{ziemann2020parameter} designs Gaussian-based privacy filters for linear Gaussian systems to prevent inference of system dynamics. However, these methods can be computationally inefficient in high-dimensional spaces. To address this, \citet{nekouei2022model} proposes a privacy filter using nonlinear transformations that retain the measurement distribution family, though it doesn't guarantee optimality in privacy metrics. More recently, \citet{weng2023optimal} introduces an approach that formulates privacy protection as a dynamic optimization problem, improving parameter privacy in dynamic systems through adaptive, state-dependent strategies.

Nevertheless, existing methods face two major limitations. First, the frameworks in \cite{erdemir2020privacy,weng2023optimal} can be inefficient in handling systems with high-dimensional or continuous state space, as they involve enumeration of past observations, which becomes exponentially complex with increasing state dimensions. Second, these methods generally assume that the data provider’s system operations are independent of its data-sharing mechanisms, implying that distorted data has no impact on the provider's system. However, in reality, data sharing often occurs between coupled systems, where the data provider’s operations are influenced by the decisions of the data user, which depend on the shared data. For instance, mobility data shared with traffic authorities may lead to changes in traffic signal settings, affecting the mobility provider’s operations.

\emph{Statement of Contribution}. This paper addresses the aforementioned challenges by proposing an interaction-aware parameter privacy-preserving data-sharing method. We make two main contributions. First, we propose an interaction-aware privacy-preserving approach based on information theory to protect operational parameters from inference attacks in a coupled system. 
Our approach generates distorted data by minimizing a combination of (i) mutual information that quantifies the privacy leakage of sensitive parameters and (ii) the impact of distorted data on the data provider’s control performance, considering the interactions between stakeholders. Our approach can successfully balance privacy preservation with control performance. Second, we formulate the optimization problem into a Bellman equation and propose a particle filter reinforcement learning (RL)-based approach to solve the formulated optimization problem. Compared to existing RL-based methods \citep{weng2023optimal, erdemir2020privacy,zhang2022privacy}, our formulation significantly reduces history dependency and efficiently handles scenarios with continuous state space.  

\section{Problem Statement}

\label{sec:Problem Statement}

\begin{figure*}[htp!]
    \centering
    \vspace{-10pt}
    \includegraphics[width=0.8\linewidth]{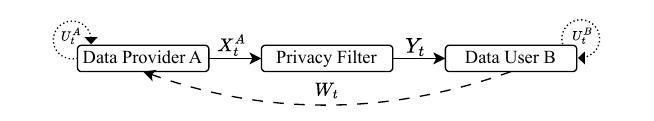}
    \caption{Interaction Between System $A$ and System $B$.}
    \vspace{-8pt}
    \label{fig:systemdynamic}
\end{figure*}

We consider the data sharing between a data provider (denoted by $A$) and a data user (denoted by $B$) over a discretized time period $\mathcal{T} = \{0,1,\cdots, T\}$ with time interval $\Delta T$. The operations of both parties are modeled as discrete-time Markov systems. Specifically, the states of the data provider and the data user at time step $t$ are denoted by $X_t^A \in \mathbb{R}^{N_A} $ and $X_t^B \in \mathbb{R}^{N_B} $, respectively. 
The state transition of $A$ follows $p(X_{t+1}^A \mid X_t^A, U_t^A)$, where $U_t^A \in \mathbb{R}^{M_A}$ is an action determined by the policy $p(U_t^A \mid \Theta, X_t^A, W_t)$. Here, $\Theta \in R^{\theta}$ is a sensitive, time-invariant parameter treated as business secrets, and $W_t$ is the external input from $B$ as a function of $X_t^B$. The state transition of $B$ follows $p(X_{t+1}^B \mid X_t^B, U_t^B)$, where the action $U_t^B \in \mathbb{R}^{M_B}$ is determined by the policy $p(U_t^B \mid X_t^B, Y_t)$. 
To support $B$'s decision-making, $A$ agrees to share information about its state $X_t^A$. Nevertheless, to protect the sensitive parameter $\Theta$, $A$ employs a data-sharing policy $\pi_t$ to generate a slightly distorted version of the data $Y_t$. Such distortion safeguards $\Theta$ from being inferred by $B$. Let $\Pi$ denote the set of feasible policies. The interaction dynamics are illustrated in Fig.~\ref{fig:systemdynamic}.

We make the following assumptions regarding the knowledge of the data provider and the data user. First, all conditional probabilities represented by $p(\cdot \mid \cdot)$ are assumed to be public knowledge. This allows us to simplify the system transitions of $A$ and $B$ to $p(X_{t+1}^A \mid \Theta, X_t^A, X_t^B)$ and $p(X_{t+1}^B \mid X_t^B, Y_t)$, respectively. Second, data user $B$ is assumed to have prior knowledge about $\Theta$, represented by a prior distribution $p(\Theta)$, but does not know its true value $\theta^*$. Three, we conservatively
assume that data user $B$ has full knowledge of the data-sharing policy $\pi_t$ at any time $t$. If $B$ does not know $\pi_t$, the privacy protection of our mechanism is only further enhanced.

We aim to design a data-sharing policy $\pi_t$ at any time $t$ for data provider $A$ that satisfies three key requirements: (i) protecting the privacy of the sensitive parameter $\Theta$, (ii) maintaining control performance of data provider $A$ , and (iii) ensuring the usability of the shared data. We make the following remarks regarding these requirements. First, continuous access to shared data allows data user $B$ to update its belief $\beta_t(\Theta)$ on $\Theta$ via Bayesian inference, so the policy must prevent $B$ from inferring the exact value of $\Theta$ over time. Second, due to the interactions between $A$ and $B$, the distorted data $Y_t$ generated by $A$ will influence the future states of $B$ and, in turn, the future states of $A$. Therefore, the data-sharing policy $\pi$ can affect the system operations of the data provider $A$.  Three, to ensure meaningful collaboration, the shared data must remain useful to $B$, requiring that the deviation between $Y_t$ and $X_t^A$ be upper bounded.

To satisfy these requirements, we formulate the data-sharing policy design into the following optimization problem\vspace{-0.8em}
\begin{subequations}    
\begin{align}
\min_{{\boldsymbol{\pi}} = \{\pi_t\}^T_{t=1}} ~ & \rho I^{\boldsymbol{\pi}}(\Theta;Y_{1:T},X^B_{1:T}) + \sum_{t=1}^T \mathbb{E}^{\boldsymbol{\pi}}\left[r(\theta^*, X_t^A, X_{t+1}^A, Y_t)\right]\label{eq:obj_org} \\
\text{s.t.}\quad   &\mathbb{E}^{\pi_t} [ d(X_t^A, Y_t) ] \leq \hat{D},\quad t = 1, 2,\cdots,T,\label{eq:con_org}
\end{align}\label{eq:original optimization}
\end{subequations}
where the objective function \eqref{eq:obj_org} includes a combination of a privacy measure and system performance measure, weighted by $\rho$. The first term is the privacy measure represented by the mutual information (MI) between the sensitive parameter $\Theta$ and $(Y_{1:T},X^B_{1:T})$, i.e., the data sequence shared by $A$ and the state data sequence of $B$. The MI quantifies the amount of information about $\Theta$ contained in $(Y_{1:T},X^B_{1:T})$. 
The second term represents the system cost over the data-sharing period, where $r(\theta^*, X_t^A, X_{t+1}^A, Y_t)$ quantifies the impact of the distorted data $Y_t$ when transitioning from state $X_t^A$ to $X_{t+1}^A$, considering the true parameter $\theta^*$. Constraint \eqref{eq:con_org} ensures that at each time step $t$, the expected distortion $d(X_t^A, Y_t)$ between the true state $X_t^A$ and the distorted state $Y_t$ under policy $\pi_t$ is upper-bounded by $\hat{D}$.

It is challenging to compute $I(\Theta;Y_{1:T}, X^B_{1:T})$ due to the high dimensionality of variables and the complexity of estimating joint distributions of $(Y_{1:T}, X^B_{1:T})$ across the entire data-sharing period. However, as proved in Appendix~\ref{appendix: MI simplification}, $I(\Theta;Y_{1:T}, X^B_{1:T})$ can be simplified as \eqref{eq:MI2} by applying the chain rule of MI and considering the dependencies between variables: 
\begin{align}
    I^{\boldsymbol{\pi}}(\Theta;Y_{1:T},X^B_{1:T}) = \sum_{t=1}^{T}I^{\boldsymbol{\pi}}(\Theta; Y_t \mid Y_{1:t-1}, X^B_{1:t}). \label{eq:MI2}
\end{align}

\section{Characterization of Interaction-Aware Privacy-Preserving Data-Sharing Policy}
\label{sec: A Computationally Efficient Privacy-Preserving Policy}
 
In this section, we devise an interaction-aware privacy-preserving data-sharing policy of data provider $A$ by solving optimization problem~\eqref{eq:original optimization}. 
A general form of the data-sharing policy is given by $\pi_t(y_t \mid \Theta, X^A_{1:t}, Y_{1:t-1}, X^B_{1:t})$, which generates the distorted data $Y_t$ given the parameter $\Theta$, histories of observations $X^A_{1:t}$ and $X^B_{1:t}$, and the history of shared data $Y_{1:t-1}$. Such a general form exploits all information available at time $t$, making the solution of Eq.~\eqref{eq:original optimization} intractable due to the curse of dimensionality. To address this issue, we equivalently simplify this policy such that it only depends on the current system state, the sensitive parameter, and the belief state. 

We first show that the optimal policy for optimization problem \eqref{eq:original optimization} can be simplified to $\pi^s_t(Y_t \mid \Theta, X^A_t, X^B_{1:t}, Y_{1:t-1})$, which eliminates dependency on $A$'s past states. Let $\boldsymbol{\pi^s} = \{\pi^s_t\}_{t=1}^T$ be the sequence of simplified policies over the entire time horizon and $\boldsymbol{\Pi^s}$ the set of all feasible simplified policies. Such a simplification is formalized in Theorem~\ref{th:simplified policy theorem} and proved in Appendix~\ref{appendix:simplified policy theorem} in the supplemental material.
\begin{theorem} \label{th:simplified policy theorem}
There exists a sequence of policy $\boldsymbol{\pi^{s,*}} \in \boldsymbol{\Pi^s}$ such that $\boldsymbol{\pi^{s,*}}$ is an optimal solution to the optimization problem \eqref{eq:original optimization}. Specifically, let $L(\cdot)$ denote the optimization problem \eqref{eq:original optimization} and $\boldsymbol{\pi^*}$ be the optimal sequence of policies, we have 

\noindent (i) $L(\boldsymbol{\pi^{s,*}})=L(\boldsymbol{\pi^*})$,

\noindent (ii) \(\mathbb{E}^{\pi^{s,*}_t} [ d(X_t^A, Y_t) ] \leq \hat{D},\quad i= 1,2,\cdots,T\).  
\end{theorem}
 
According to Theorem~\ref{th:simplified policy theorem}, the optimal simplified policy can be obtained from the following optimization problem
\begin{align}
    \min_{\boldsymbol{\pi^s}}\sum_{t=1}^T &\Bigg(\rho I^{\boldsymbol{\pi^s}}(\Theta; Y_t\mid Y_{1:t-1},X^B_{1:t})+ \mathbb{E}^{\boldsymbol{\pi^s}}[r(\theta^*,X_t^A,X^A_{t+1},Y_t)] + \lambda\Big(\mathbb{E}^{\boldsymbol{\pi^s}}[d(X^A_t,Y_t)]-\hat{D}\Big)\Bigg)\label{eq:simplified_policy_optimization}
\end{align}
where $\lambda$ is the Lagrangian multiplier. To solve optimization problem \eqref{eq:simplified_policy_optimization}, we introduce the cost-to-go function $V_t(h_t)$, which represents the minimum expected cost starting at time $t$ given the history $h_{t} = (x^B_{1:t}, y_{1:t-1})$. This function $V_t(h_t)$ allows us to recursively express the optimization problem and solve it using dynamic programming. Specifically, the optimal cost-to-go at time $t$ is defined as
\begin{align}\label{eq:cost_to_go}
    V_t^{*}(h_{t}) = \min_{\{\pi_k^s\}_{k=t}^T}  \Bigg[ \sum_{k=t}^T &\Bigg( \rho I^{\boldsymbol{\pi^s}}(\Theta; Y_k \mid Y_{1:k-1}, X^B_{1:k})+ \mathbb{E}^{\boldsymbol{\pi^s}}[r(X_k^A, X_{k+1}^A, Y_k, \theta^*)] \notag\\ &\quad + \lambda\Big(\mathbb{E}^{\boldsymbol{\pi^s}}[d(X^A_k,Y_k)]-\hat{D}\Big)\Bigg)\Bigm| h_{t} \Bigg]
\end{align}
with terminal condition $V^{*}_{T+1}(\cdot) = 0$. Accordingly, the Bellman optimality equation of \eqref{eq:simplified_policy_optimization} can be represented as 
\begin{align}
\label{eq:Bellman original}
    V^{*}_t(h_{t}) = \min_{\pi^s_t(Y^t\mid\Theta,X^A_t,h_t)} \left[C_t(h_{t},\pi^s_t(Y_t\mid\Theta,X^A_t,h_t)) + \mathbb{E}\left(V^{*}_{t+1}(h_{t+1}\mid h_{t})\right)\right],
\end{align}
where $C_t(h_{t}, \pi^s_t(Y^t\mid\Theta,X^A_t,h_t))$ represents the immediate cost when employing data-sharing policy set $\pi^s_t(Y^t\mid\Theta,X^A_t,h_t)$, with detailed formulation in Appendix~\ref{appendix:proof for bellman equiv}.

Although the simplified policy $\pi^s$ reduces complexity by eliminating dependence on $A$'s past states $X^A_{1:t-1}$, it still relies on the histories $h_{t}$. This dependence complicates the optimization problem as the time horizon $T$ increases. To overcome this challenge and efficiently generate the policy set $\pi^s$ while utilizing information from the histories $h_t$, we propose an alternative formulation. We encode the histories into a belief state $\beta_t$, which condenses all past observational information into a probability distribution over the possible current system states $\beta_{t}(\Theta, X_t^A) = p(\Theta, X_t^A \mid h_{t})$. The belief state is updated recursively using Bayes' rule whenever new observations become available. Specifically, upon receiving the distorted data $y_t$ at time $t$, the data user updates the belief state from $\beta_t$ to $\beta_{t+1}$ using the following formula:
\begin{align}
\label{eq:belief_update}
    & \beta_{t+1}(\Theta, X_{t+1}^A) =\frac{\int_{x^A_t}p(X^A_{t+1}\mid \Theta,x^A_t,x_t^B)a(y_t\mid\Theta,x^A_t)\beta_{t}(\Theta, x_{t}^A)dx^A_t}{\int_{x^A_t,x^A_{t+1},\theta}p(x^A_{t+1}\mid \theta,x^A_t,x_t^B)a(y_t\mid\theta,x^A_t)\beta_{t}(\theta, x_{t}^A)d\theta dx^A_{t+1}dx^A_t},
\end{align}
Where $a(y_t\mid \Theta,x^A_t)= \pi^s_t(Y_{t}\mid \Theta,x^A_t,X^B_{1:t},Y_{1:t-1})$.\

For a detailed derivation of this update rule, please refer to the supplementary material, Appendix~\ref{appendix:update for belief}.

According to  \eqref{eq:belief_update}, we can see the belief state $\beta_t$ relies solely on the observations $Y_t$, policy set $\pi^s_t$ and the prior belief $\beta_{t}$. Then, similar to the idea in~\cite{weng2023optimal}, we establish Lemma~\ref{le:bellman equiv} to show that the optimal cost-to-go function, derived using the belief state $\beta_t$, is equivalent to that derived using the full history $h_t$. The proof is in the supplementary material (Appendix~\ref{appendix:proof for bellman equiv}).

\begin{lemma}\label{le:bellman equiv}
At any time step $t$, given histories $h_t$, the optimal cost-to-go function $V_t^*(h_t)$ depends only on the belief state $\beta_t$, i.e., $V_t^*(h_t) = V_t^*(\beta_t)$
\end{lemma}

Based on Lemma~\ref{le:bellman equiv}, given histories $h_t$, the data-sharing policy $\pi^s_t(Y_t \mid \Theta, X^A_t, h_t)$ defines a Markov kernel $\mathcal{K}_t$ that maps the state variables $(\Theta, X^A_t)$ to a probability distribution over $Y_t$. For any measurable spaces $(\Theta, X^A_t)$ and $Y_t$, $\mathcal{K}_t$ assigns a conditional distribution to $Y_t$ based on $\pi^s_t$. Accordingly, the Bellman optimality equation can be expressed as:
\begin{align}\label{eq:Bellman kernel}
    V_t^*(\beta_t) = \min_{\mathcal{K}_t} \big[ C_t(\beta_t, \mathcal{K}_t) + \mathbb{E}[V_{t+1}^*(\beta_{t+1}) \mid h_t] \big], 
\end{align}
which is derived from \eqref{eq:Bellman original} by replacing $h_t$ by $\beta_t$ and $\pi_t^s$ by $\mathcal{K}_t$. 

We deduce from \eqref{eq:Bellman kernel} that the optimal Markov kernel $\mathcal{K}^*_t$ depends only on the belief state $\beta_t$, i.e., $\mathcal{K}^*_t = \arg \min V_t^*(\beta_t)$. Therefore, we consider $\mathcal{K}_t$ as a function of $\beta_t$, denoted by $\mathcal{K}_t^{\beta_t}$. Consequently, the data-sharing policy can be determined solely by $\beta_t$, $\Theta$, and $X^A_t$, summarized in Theorem~\ref{th:optimal policy}. 

\begin{theorem}\label{th:optimal policy}
Let $\mathcal{K}^{\beta_t,*}_t$ denote the solution to Eq.~\eqref{eq:Bellman kernel}. Given histories $h_t$, the optimal data distortion policy satisfies:
\begin{align}
\vspace{-10pt}
    \pi^{s,*}_t(Y_t \mid \Theta, X^A_t, h_t) = \mathcal{K}^{\beta_t,*}_t,
\end{align}
where $\beta_t$ is the belief state associated with $h_t$.
\end{theorem}

Theorem~\ref{th:optimal policy} simplifies the data-sharing policy by demonstrating that it is fully determined by the current state $(\Theta, X^A_t)$ and the belief state $\beta_t$. Therefore, for brevity, we use $\mathcal{K}^{\beta_t,*}_t(Y_t\mid\Theta, X^A_t)$ to represent the policy $\pi^{s,*}_t(Y_t\mid \Theta, X^A_t,h_t)$ in the rest of the article.

\section{Particle Filter Reinforcement Learning-Based Solution Method}
\label{sec: Utility-Aware Privacy-Preserving Data Sharing Framework}

\begin{figure}[t]
    \centering
    \vspace{-5pt}
    \includegraphics[width=1.1 \linewidth]{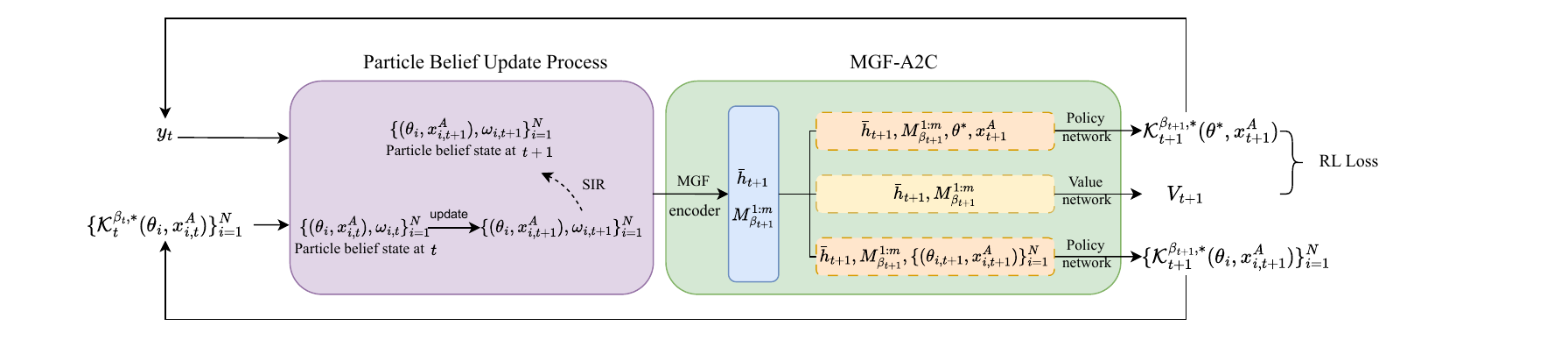}
    \caption{Overview of the proposed privacy-preserving data-sharing framework.}
    \vspace{-10pt}
    \label{fig:Architecture}
\end{figure}

In this section, we devise a solution method to solve the interaction-aware privacy-preserving data-sharing policy characterized in Section~\ref{sec: A Computationally Efficient Privacy-Preserving Policy}. Unlike existing works \citep{weng2023optimal,erdemir2020privacy} that focus on low-dimensional discrete state space, our method combines particle filters (PF) and reinforcement learning (RL) to handle continuous state space with relatively high dimensions. As shown in Fig.~\ref{fig:Architecture}, we employ a PF to characterize the evolution of the belief state $\beta_t$, which is then processed by a Particle Advantage Actor-Critic (A2C) algorithm with a Moment-Generating Function (MGF) encoder, hereafter named MGF-A2C, to generate the data-sharing policy. 

\subsection{Particle formulation of belief state}\label{subsec:PF}

The belief state $\beta_t$ is characterized by the posterior distribution of $(\Theta, X^A_t)$, which is challenging to compute for continuous $(\Theta, X^A_t)$. To address this challenge, we describe the evolution of the belief state $\beta_t$, represented by~\eqref{eq:belief_update}, using a PF, which approximates the posterior distribution of $(\Theta, X^A_t)$ with weighted sampled particles. As the system evolves, particles move according to the dynamic model, and their weights are updated based on how well they match new observations.

Specifically, we use a set of $N$ weighted particles $\{(\theta_{i}, x^A_{i,t}), \omega_{i,t}\}_{i=1}^N$ to estimate $\beta_t$. Each particle $(\theta_{i}, x^A_{i,t})$ represents a sampled possible state from $p(\Theta, X^A_t \mid Y_{1:t}, X^B_{1:t})$ at time $t$, and $\omega_{i,t}$ is the associated weight, reflecting the importance of this particle, with the condition $\sum_{i=1}^N \omega_{i,t} = 1 $. Thus, the estimated belief state $\hat{\beta}_{t}$ is given by the set $\hat{\beta}_{t} = \{(\theta_{i}, x^A_{i,t}), \omega_{i,t}\}_{i=1}^N$. The weight update process can be formalized in Theorem~\ref{thm:weight_update}. The proof is in Appendix~\ref{appendix:weight_update_derivation} in the supplementary material. 

\begin{theorem}\label{thm:weight_update} In the PF approximation of the belief state $\beta_t(\Theta, X^A_t)$, the weights $\omega_{i,t}$ can be updated recursively using the observation likelihood as 
\begin{align} \omega_{i,t} \propto \omega_{i,t-1}p(X^B_{t}\mid X^B_{t-1},y_{t-1})p(y_{t-1}\mid \theta_{i},x^A_{i,t-1},X^B_{1:t-1},Y_{1:t-2}). 
\end{align} 

\end{theorem}
These updated weights are normalized as $\omega_{i,t} = \frac{\omega_{i,t}}{\sum_{j=1}^N \omega_{j,t}}$ to ensure that they form a valid probability distribution.
Accordingly, we can easily derive Lemma~\ref{le:particle belief update}, which further simplifies Theorem~\ref{thm:weight_update} to be independent of the dynamics of system $B$. 

\begin{lemma}
\label{le:particle belief update}
    The weight update $\omega_{i,t} \propto \omega_{i,t-1}p(X^B_{t}\mid X^B_{t-1},y_{t-1})p(y_{t-1}\mid \theta_{i},x^A_{i,t-1},X^B_{1:t-1},Y_{1:t-2})$ is equivalent to $\omega_{i,t}\propto \omega_{i,t-1}p(y_{t-1}\mid \theta_{i,t-1},x^A_{i,t-1},X^B_{1:t-1},Y_{1:t-2})$.
\end{lemma}
Sequential Importance Resampling (SIR)~\citep{rubin1981bayesian} is adopted to mitigate the degeneracy phenomenon. The details can be found in Appendix~\ref{appendix: paricle formulation setting} in the supplementary material.

With this recursive weight update rule, given shared data $y_t$ (observation) and the data-sharing policy $\mathcal{K}^{\beta_t,*}_t(Y_t\mid\Theta, X^A_t)$, we can calculate $\hat{\beta}_t$ using $\{(\theta_{i},x^A_{i,t}),\omega_{i,t} \}^N_{i=1}$ with weight update process $\omega_{i,t+1} = \mathcal{K}^{\beta_t,*}_t(y_t\mid\theta_i, x^A_{i,t})\omega_{i,t}$, and the state of each particle will evolve through the system dynamics described in Section~\ref{sec:Problem Statement}.

\subsection{A2C solution algorithm with MGF encoder}
\label{subsec: A2C}
With the particle representation of the belief state, we solve the optimal policy $\mathcal{K}^{\beta_t,*}_t(Y_t\mid\Theta, X^A_{t})$ for each particle using an A2C algorithm. 
As illustrated in Fig.~\ref{fig:Architecture}, our proposed algorithm comprises three main components: an MGF encoder, an actor network, and a critic network. 
The MGF encoder extracts features from the estimated particle-based belief state $\hat{\beta}_t$. 
This is because high-precision belief estimation often requires a large number of particles, leading to high-dimensional inputs to the actor and critic networks, which can make training computationally inefficient or even unstable. Additionally, SIR introduces particle permutation, which further complicates convergence. To address these issues, we adopt the MGF method \citep{ma2020discriminative} to efficiently encode higher-order moments into a low-dimensional vector. The MGF captures all moments of a distribution, fully characterizing it while reducing dimensionality \citep{bulmer2012principles}. Details of the mapping from particle beliefs to MGF features are provided in \cite{ma2020discriminative}. The actor network takes features from MGF features $(\bar{h}_t, M_{\beta_t}^{1:m})$ (detailed definition can be found in Appendix~\ref{appendix: paricle formulation setting}), $\theta$, and $x^A_t$ as input, and outputs the parameters of a Dirichlet distribution, from which the policy $\mathcal{K}^{\beta_t}_t(Y_t\mid\theta, x^A_t)$ is sampled. The critic network takes the features from MGF as input and estimates a value function, which is used to update the actor network.

 \emph{Training Procedure}. The actor and critic networks, parameterized by $\xi$ and $\theta_{\text{critic}}$, respectively, are trained using standard A2C methods. The MGF encoder's parameters, $\phi \in \Phi$, are optimized jointly with the actor's policy loss, $\mathcal{L}_a(\xi)$, and the critic's value loss, $\mathcal{L}_c(\theta_{\text{critic}})$. This joint optimization enables the encoder to learn features that are beneficial for both action selection and value estimation. The MGF encoder’s parameters are updated as $\phi_{t+1} = \phi_t - \eta_a \nabla_{\phi} \mathcal{L}_a(\xi_t) - \eta_c \nabla_{\phi} \mathcal{L}_c(\theta_{{\text{critic}},t})$, 
where $\mathcal{L}_a(\xi_t) = -\ln(q_{\xi_t}(a_t|\beta_t)) \delta_t$ is the actor's loss with Temporal Difference (TD) error $\delta_t$, $ \mathcal{L}_c(\theta_{{\text{critic}},t}) = \delta_t^2 $ is the critic's loss, minimizing the squared TD error, and $\eta_a$ and $\eta_c$ are the learning rates for the actor and critic, respectively. This dual-gradient update scheme allows the MGF encoder to optimize feature representations that benefit from both the actor's policy improvement and the critic's value estimation.

\section{Experiments}
\label{sec:Simulation Settings and Results}
In this section, we validate our proposed interaction-aware privacy-preserving data-sharing method in a mixed-autonomy platoon control scenario~\citep{ZHOU2024104885}. 
\begin{figure}[htp!]
    \centering
    \vspace{-5pt}
    \subfigure[Mixed-autonomy platoon.]{    \includegraphics[width=7cm]{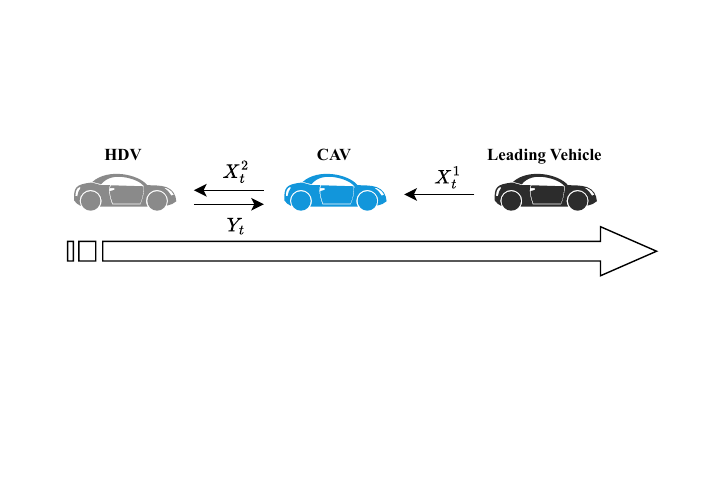}  
        \label{fig: mixed-platoon}}
    \hfill 
    \subfigure[Training reward per episode.]{        \includegraphics[width=7cm]{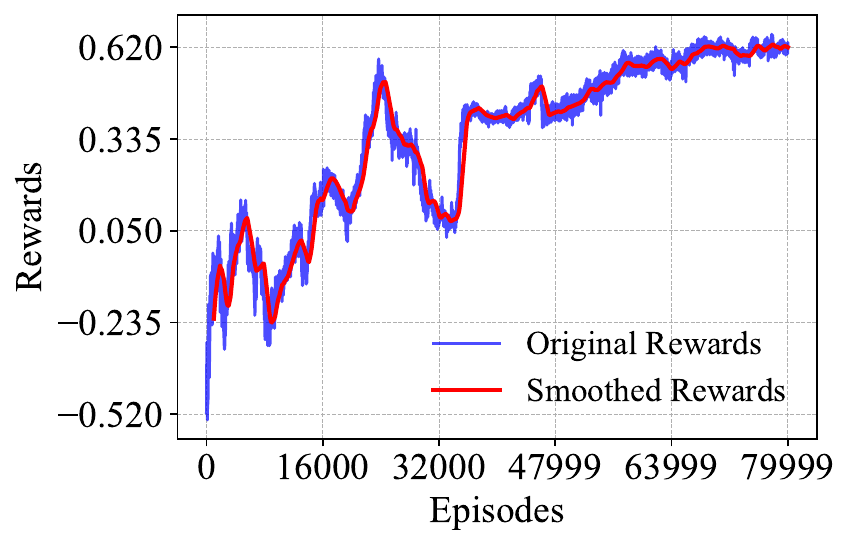} 
        \label{fig:training}
    } \caption{Experiment settings and training results}
    \vspace{-10pt}
\end{figure}

\subsection{Experimental setup}
\label{subsec: experiment setup}
Our experimental setup includes scenario settings, adversaries, training, and evaluation. For a detailed description, please refer to the supplementary material (i.e., Appendix~\ref{appendix: Experiment}).

\noindent \textbf{Scenario settings}. 
As shown in Fig.~\ref{fig: mixed-platoon}, we consider a mixed-autonomy platoon comprising a leading vehicle, a connected and automated vehicle (CAV) in the middle (data user $B$), and a following human-driven vehicle (HDV) at the tail (data provider $A$). The CAV gathers state information (speed and spacing) from all vehicles through vehicle-to-vehicle communications and computes its optimal control inputs accordingly. Therefore, this historical data shared by HDV could be used by adversaries to infer driving behavior parameters, which are the sensitive parameters we aim to protect. The modeling of these vehicles is as follows. Similar to~\citet{zhou2024enhancing}, the leading vehicle seeks to maintain its speed around a desired speed, and the speed tracking error follows a Gaussian distribution with a mean of $0$ and a variance of $0.1$. The details of system dynamics of the CAV and HDV can be referred to \citet{wang2021leading,zhou2024enhancing}. The CAV control action $u(t)$ is determined using a distributed linear controller following \citet{wang2021leading}. The following HDV’s car-following behavior is parameterized by the sensitive parameter $ \Theta$, which is an intrinsic property of a human driver.

\noindent \textbf{Adversaries}. Adversaries seek to infer the sensitive car-following parameters from the HDV using the system state of the CAV and the shared data of the HDV. Specifically, we assume that the adversaries can use Bayesian Inference (BI) and Recursive Least Squares (RLS)~\citep{Haykin:2002} to infer the sensitive parameters. It is noted that BI is modeled as a PF that updates beliefs about system parameters using an observation model with Gaussian noise. 

\noindent\textbf{Training and evaluation}: The proposed MGF-A2C algorithm is applied to learn the optimal data-sharing policy. The network architecture and hyperparameters are given in Appendix~\ref{appendix: Experiment} in the supplementary material. The data sharing is performed at a frequency of $5$\,Hz. The training process considers data sharing of $40$\,s, whereby the belief states are characterized by $324$ particles ($4$ for $\Theta$ and $81$ for $X^A_1$), which are initially uniformly distributed within the parameter and state space with equal weight. The training rewards per episode are presented in Fig.~\ref{fig:training}, showing convergence after 60,000 episodes. 
To evaluate the trained model, the simulation was run for 1200 time steps (equivalent to 10 minutes), utilizing 5184 particles to ensure comprehensive assessment. We adopt the evaluation settings with longer and finer particle representation to more accurately measure the privacy leakage. For each value of $\Theta$, experiments were conducted 10 times to evaluate privacy (i.e., the accuracy of  RLS-estimated parameters) and control performance (i.e., fuel consumption).

\subsection{Privacy-preserving performance}
\label{subsec: Privacy-Preserving Performance Analysis}
We evaluate the privacy-preserving performance of our proposed data-sharing algorithm (F) by comparing it against the benchmark policy that shares true data (R) without privacy protection. 

\begin{figure*}[t]
    \centering
    \parbox{0.45\linewidth}{%
        \centering
        \includegraphics[width=\linewidth]{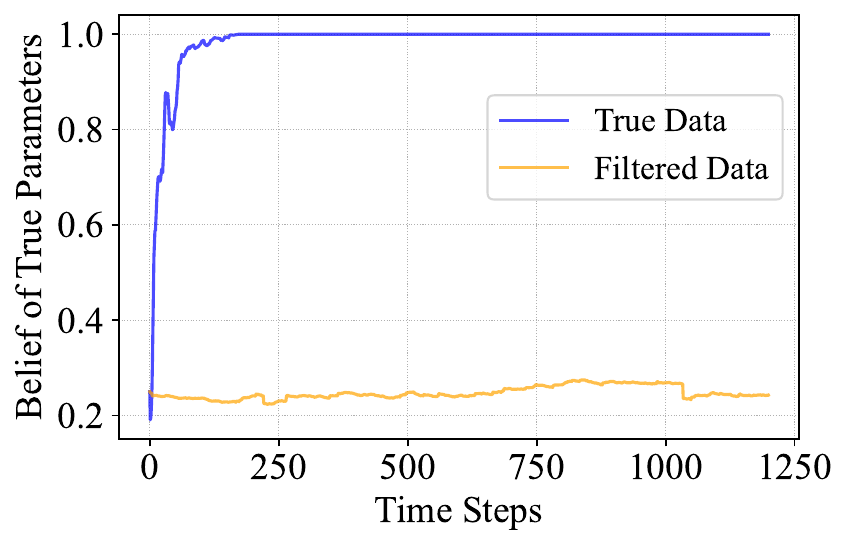}
        \\ $\theta_1 = (0.4, 0.5)$ 
    }\hspace{10pt}
    \parbox{0.45\linewidth}{%
        \centering
        \includegraphics[width=\linewidth]{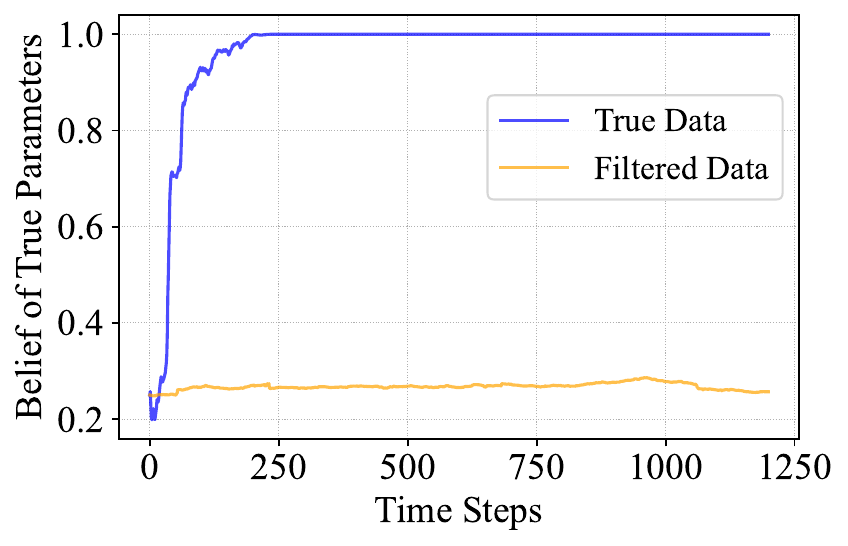}
        \\ $\theta_2 = (0.7, 0.8)$ 
    }

    \vspace{10pt} 

    \parbox{0.45\linewidth}{%
        \centering
        \includegraphics[width=\linewidth]{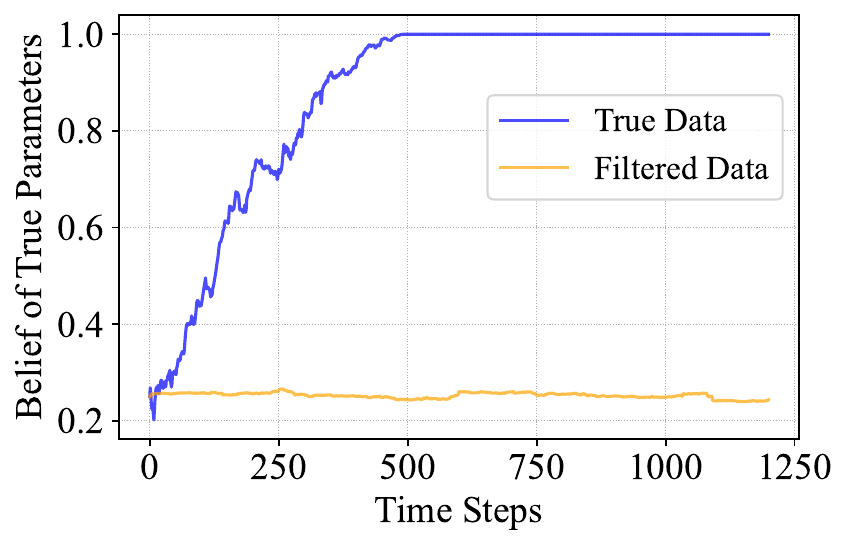}
        \\$\theta_3 = (1, 1.1)$ 
    }\hspace{10pt}
    \parbox{0.45\linewidth}{%
        \centering
        \includegraphics[width=\linewidth]{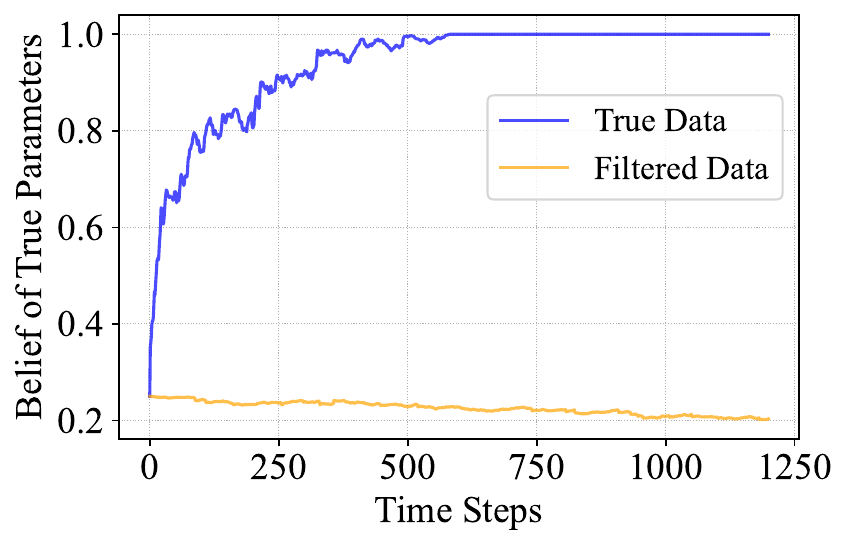}
        \\ $\theta_4 = (1.3, 1.4)$ 
    }

    \vspace{10pt}
    \caption{Comparison of Bayesian attacker's belief over time for different values of $\theta$. Each figure shows the evolution of the attacker's belief when observing true data versus distorted (filtered) data.}
    \label{fig:parameter-comparison}
    \vspace{-10pt}
\end{figure*}

\noindent\textbf{Inference Attack using BI}: BI is applied to quantify the privacy leakage represented by the belief of true parameters over time. The resulting evolution of belief is shown in Fig.~\ref{fig:parameter-comparison} for different values of true theta. As we can see from Figure~\ref{fig:parameter-comparison}, the posterior probability resulting from our data-sharing algorithm remains largely similar to the prior probability which is uniform (i.e.,$p(\theta^*)$ is around 0.25). This indicates that the attacker is unable to significantly improve their belief about the $\theta^*$ from the distorted data sequence, demonstrating the effectiveness of the privacy filter in preventing sensitive information leakage. In contrast, when BI estimates $\theta^*$ only based on the true data sequence it observed, its belief about $\theta^*$ quickly converges to a value close to 1, showing that they can accurately infer the true parameters.

\noindent\textbf{Inference Attack using RLS}: We compare the parameters estimated by RLS using the log-transformed Root Mean Square Error (RMSE, $\sigma_e$) between $\theta^*$ and the estimated parameters. This metric, referred to as the success rate (SR), maps the RMSE to a range between 0 and 1, with values closer to 1 indicating a higher success rate of the $\theta^*$ estimation. The results are illustrated in Table.~\ref{Tb: Combined Table}. We can clearly observe that the benchmark policy yields a small RMSE of the estimated parameter, which means the RLS can accurately estimate the exact value of driving behavior. In contrast, the RMSE of the estimated parameter from the distorted data is larger, indicating that our framework effectively reduces the RLS’s ability to accurately estimate the exact values of $\Theta$.

These results confirm that our framework effectively safeguards driving behavior information, regardless of the attack method used. Both BI and RLS-based attacks show that our method consistently limits the attacker’s ability to deduce the true value of $\Theta$, ensuring privacy protection across various conditions.

\begin{table*}[ht]
    \renewcommand\arraystretch{1.3} 
    \small
    \centering
    \caption{Comparison of Parameter Estimation Accuracy and Fuel Consumption}
    \label{Tb: Combined Table}
    \begin{tabular}{>{\centering\arraybackslash}p{1.1cm} >{\centering\arraybackslash}p{1.2cm} >{\centering\arraybackslash}p{1.2cm} >{\centering\arraybackslash}p{1.2cm} >{\centering\arraybackslash}p{1.2cm} >{\centering\arraybackslash}p{1.2cm} >{\centering\arraybackslash}p{1.2cm} >{\centering\arraybackslash}p{1.2cm}}
        \toprule
        \multirow{2}{*}{\textbf{$\Theta$}} 
        & \multicolumn{4}{c}{\textbf{Privacy Measures for Parameters}} 
        & \multicolumn{3}{c}{\textbf{Fuel Consumption}} \\
        \cmidrule(lr){2-5} \cmidrule(lr){6-8}
        & \textbf{$\sigma_e$ (R)} & \textbf{SR (R)} 
        & \textbf{$\sigma_e$ (F)} & \textbf{SR (F)} 
        & \textbf{R} & \textbf{F} & \textbf{$\Delta\%$} \\
        \midrule
        $\theta_1$ & 0.0019 & 0.9981 & 2.928 & 0.0535 & 1.254 & 1.273 & 1.54 \\
        $\theta_2$ & 0.0075 & 0.9925 & 3.581 & 0.0278 & 1.260 & 1.291 & 2.47 \\
        $\theta_3$ & 0.0024 & 0.9976 & 2.280 & 0.1023 & 1.268 & 1.307 & 3.14 \\
        $\theta_4$ & 0.0030 & 0.9970 & 2.454 & 0.0859 & 1.283 & 1.322 & 3.03 \\
        \bottomrule
    \end{tabular}
    \vspace{-10pt}
\end{table*}

\subsection{Tradeoff between privacy and control performance}
\label{subsec:Privacy-Utility Trade-off Analysis}
To assess the impact of the proposed data-sharing policy, we compare the HDV's fuel consumption performance (mL/s) before and after applying the filter and calculate the difference $\Delta$. The experimental results are summarized in Table~\ref{Tb: Combined Table}. The results indicate that for different driving behaviors $\Theta$, our data-sharing policy results in only a $1.5\%$ to $3\%$ increase in fuel consumption compared to directly sharing real state data, which is negligible. This demonstrates that our data-sharing policy effectively protects sensitive information with minimum impact on the control performance.

\section{Conclusion}
 
In this paper, we propose an interaction-aware privacy-preserving data-sharing approach to facilitate data sharing between coupled systems with continuous state space, where the mutual impact between systems introduces additional complexity. Our approach generates distorted data by minimizing a combination of (i) mutual information that quantifies the privacy leakage of sensitive parameters and (ii) the impact of distorted data on the data provider’s control performance, considering the interactions between stakeholders. The optimization problem is formulated into a Bellman equation and solved by a particle filter reinforcement learning (RL)--based approach. Case studies on mixed-autonomy platoon control demonstrate the ability of our approach to protect sensitive parameters, such as HDV driving behaviors, while ensuring minimal impact on control performance. Future work includes extending the proposed approach to the sharing of high-dimensional operational data (e.g., vehicle trajectories). It would also be interesting to investigate the data-sharing incentives between multiple stakeholders using game-theoretic approaches.

\acks{This research was supported by the Singapore Ministry of Education (MOE) under its Academic Research Fund Tier 2 (A-8003064-00-00).}

\bibliography{ref.bib}
\appendix
\newpage
\renewcommand{\thesubsection}{\thesection.\arabic{subsection}} 
\section{Proof for Simplification of MI in Section~\ref{sec:Problem Statement}}
\label{appendix: MI simplification}
Based on the chain rules of conditional mutual information, we have:
\begin{align}
    I^{\boldsymbol{\pi}}(\Theta;Y_{1:T},X^B_{1:T}) & = \sum_{t=1}^T I^{\boldsymbol{\pi}}(\Theta;Y_t,X^B_{t}\mid Y_{1:t-1},X_{1:t-1}^B) \nonumber\\
    &= \sum_{t=1}^T \Big(I^{\boldsymbol{\pi}}(\Theta; Y_t\mid Y_{1:t-1},X^B_{1:t}) + I^{\boldsymbol{\pi}}(\Theta; X^B_t\mid Y_{1:t-1},X^B_{1:t-1})\Big) \label{eq:MI1}
\end{align}

Notice that by the system dynamics of data user $B$, the impact of $\Theta$ on $X^B_t$ is mediated through $(X^B_{1:t-1}, Y_{1:t-1})$. Therefore, $X^B_t$ is independent of $\Theta$ given $(X^B_{1:t-1}, Y_{1:t-1})$. Hence, $I^{\boldsymbol{\pi}}(\Theta; X^B_t \mid Y_{1:t-1}, X^B_{t-1}) = 0$. Thus, Eq.~\eqref{eq:MI1} can be further simplified as
\begin{align}
    I^{\boldsymbol{\pi}}(\Theta;Y_{1:T},X^B_{1:T}) = \sum_{t=1}^{T}I^{\boldsymbol{\pi}}(\Theta; Y_t \mid Y_{1:t-1}, X^B_{1:t}).
\end{align}

\section{Proofs of  Sec.~\ref{sec: A Computationally Efficient Privacy-Preserving Policy}}

\subsection{Proof for Theorem~\ref{th:simplified policy theorem}}
\label{appendix:simplified policy theorem}
To prove Theorem~\ref{th:simplified policy theorem}, we first show 
\begin{align}
p^{\pi}(\Theta,X^A_{t},X^A_{t+1},X^B_{1:t},Y_{1:t})= p^{\pi_s}(\Theta,X^A_{t},X^A_{t+1},X^B_{1:t},Y_{1:t}). \label{eq:full prob}
\end{align}

To this end, let us further decompose $ p^{\pi}(\Theta,X^A_{t},X^A_{t+1},X^B_{1:t},Y_{1:t})$ as:
\begin{align}
     p^{\pi}(\Theta,X^A_{t},X^A_{t+1},X^B_{1:t},Y_{1:t}) &= p(X^A_{t+1}|\Theta,X^A_t,X_t^B,Y_{1:t})p^{\pi}(\Theta, X^A_t,X^B_{1:t},Y_{1:t})\notag\\
    &=p(X^A_{t+1}|\Theta,X^A_t,X_t^B)\pi_t(Y_t|\Theta,X^A_t,X^B_{1:t},Y_{1:t-1})p^{\pi}(\Theta,X^A_t,X^B_{1:t},Y_{1:t-1})\notag\\
    &=p(X^A_{t+1}|\Theta,X^A_t,X_t^B)\pi_t(Y_t|\Theta,X^A_t,X^B_{1:t},Y_{1:t-1})p(X_t^B|X_{t-1}^B,Y_{t-1})\notag\\&\quad p^{\pi}(\Theta,X^A_t,X^B_{1:t-1},Y_{1:t-1})\notag\\
    & = p(X^A_{t+1}|\Theta,X^A_t,X_t^B)\pi_t(Y_t|\Theta,X^A_t,X^B_{1:t},Y_{1:t-1})p(X_t^B|X_{t-1}^B,Y_{t-1})\notag\\&\quad \int_{x^A_{t-1}}p^{\pi}(\Theta,X^A_{t-1},x^A_t,X^B_{1:t-1},Y_{1:t-1})dx^A_{t-1}.\label{eq:decomposion}
\end{align}
where the second equation follows the Markov property of system dynamics of data provider $A$.

For any $\pi_t\in \Pi$, we can choose a simplified policy $\pi^s_t \in \Pi_s$  such that:
\begin{align}
    & \pi^s_t(Y_t|\Theta,X^A_t,X^B_{1:t},Y_{1:t-1})= \pi_{t,Y_t|\Theta,X^A_t,X^B_{1:t},Y_{1:t-1}}(Y_t|\Theta,X^A_t,X^B_{1:t},Y_{1:t-1}).
\end{align}

We then use induction to show \eqref{eq:full prob}. For $t=1$, we have
\begin{align}
p^{\pi}(\Theta,X^A_{1},X^A_{2},X^B_{1},Y_{1}) = p(X^A_{2}\mid 
\Theta, X^A_{1},X^B_{1}) \pi_1(Y_1\mid \Theta, X^A_{1},X^B_{1})p(\Theta, X^A_{1},X^B_{1})
\end{align}
Apparently, we have $\pi_1(Y_1\mid \Theta, X^A_{1},X^B_{1}) = \pi^s_1(Y_1\mid \Theta, X^A_{1},X^B_{1})$. Thus, the induction can be initialized as follows: 
\begin{align}
    p^{\pi}(\Theta,X^A_{1},X^A_{2},X^B_{1},Y_{1}) = p^{\pi^s}(\Theta,X^A_{1},X^A_{2},X^B_{1},Y_{1})
\end{align}
Then, assuming $
    p^{\pi}(\Theta,X^A_{t-1},X^A_t,X^B_{1:t-1},Y_{1:t-1})= p^{\pi_s}(\Theta,X^A_{t-1},X^A_t,X^B_{1:t-1},Y_{1:t-1})$, \eqref{eq:full prob} is a natural conclusion of  \eqref{eq:decomposion}. 

With \eqref{eq:full prob}, we show the equivalence of MI. Specifically, we have 
\begin{align}
    p^{\pi}(\Theta,X^B_{1:t},Y_{1:t})& =\int_{x^A_t,x^A_{t+1}}p^{\pi}(\Theta,x^A_t,x^A_{t+1},X^B_{1:t},Y_{1:t})dx^A_tdx^A_{t+1}\notag\\
    &= \int_{x^A_t,x^A_{t+1}}p^{\pi^s}(\Theta,x^A_t,x^A_{t+1},X^B_{1:t},Y_{1:t})dx^A_tdx^A_{t+1}= p^{\pi^s}(\Theta,X^B_{1:t},Y_{1:t}), \label{eq:simplification_1}
\end{align}
which leads to $I^{\pi}(\theta; Y_{1:t}, X^B_{1:t}) = I^{\pi_s}(\theta; Y_{1:t}, X^B_{1:t})$ by the definition of MI. 

We then show the equivalence of control performance under policies $\pi$ and $\pi_s$, i.e., $\mathbb{E}^{\boldsymbol{\pi}}[r(\theta^*,X^A_t,X^A_{t+1},Y_t)]$. Note that 
\begin{align}
     p^{\pi}(\theta^*,X^A_t,X^A_{t+1},Y_t) &= \int_{x^B_{1:t}}\int_{y_{1:t-1}}p^{\pi}(\theta^*,X^A_t,X^A_{t+1},X^B_t,x^B_{1:t},Y_t,y_{1:t-1})dy_{1:t-1}dx^B_{1:t}\notag\\
     & = \int_{x^B_{1:t}}\int_{y_{1:t-1}}p^{\pi_s}(\theta^*,X^A_t,X^A_{t+1},X^B_t,x^B_{1:t},Y_t,y_{1:t-1})dy_{1:t-1}dx^B_{1:t}\notag\\
     & = p^{\pi^s}(\theta^*,X^A_t,X^A_{t+1},Y_t), 
\end{align}
where the second equation results from \eqref{eq:full prob}. Then, the equivalence of control performance can be proved by the definition of expectation.

Similarly, we only need to prove $p^{\pi}(X^A_t,Y_t) = p^{\pi^s}(X^A_t,Y_t)$ for the equivalence of data deviation constraint under policies $\pi$ and $\pi_s$. This can be derived from \eqref{eq:simplification_1} by calculating the marginal probability on both sides of the equation.

This completes the proof.  

\subsection{Derivation of Belief State Update}
\label{appendix:update for belief}
The belief state update is derived using Bayes' rule. Starting with the posterior probability $\beta_{t+1}(\Theta,X^A_{t+1})$, we have
\begin{align}
    & \beta_{t+1}(\Theta,X^A_{t+1}) = p(\Theta,X^A_{t+1}\mid X^B_{1:t+1},Y_{1:t}) = \frac{p(\Theta,X^A_{t+1}, X_{t+1}^B, Y_{t}\mid X^B_{1:t},Y_{1:t-1})}{p(X_{t+1}^B,Y_t\mid  X^B_{1:t},Y_{1:t-1})}\notag\\
    & = \frac{\int_{x^A_t}p(\Theta,x^A_t,X^A_{t+1},X_{t+1}^B, Y_{t}\mid X^B_{1:t},Y_{1:t-1})dx^A_t}{\int_{x^A_t}\int_{x^A_{t+1}}\int_{\theta}p(\theta,x^A_{t},x^A_{t+1},X_{t+1}^B,Y_t\mid  X^B_{1:t},Y_{1:t-1})d\theta dx^A_{t+1}dx^A_t}\notag\\
    & = \frac{\int_{x^A_t}p(X^B_{t+1}\mid X^B_t,Y_t)p(\Theta,x^A_t,X^A_{t+1},Y_{t}\mid X^B_{1:t},Y_{1:t-1}) dx^A_t}{\int_{x^A_t}\int_{x^A_{t+1}}\int_{\theta}p(X_{t+1}^B\mid Y_t,X_t^B)p(\theta,x^A_{t},x^A_{t+1},Y_t\mid  X^B_{1:t},Y_{1:t-1})d\theta dx^A_{t+1}dx^A_t}\notag\\
    & = \frac{\int_{x^A_t}p(X^B_{t+1}\mid X^B_t,Y_t)p(X^A_{t+1}\mid \Theta,x^A_t,X_t^B)p(\Theta,x^A_t,Y_{t}\mid X^B_{1:t},Y_{1:t-1})dx^A_t}{\int_{x^A_t}\int_{x^A_{t+1}}\int_{\theta}p(X_{t+1}^B\mid Y_t,X_t^B)p(x^A_{t+1}\mid \theta,x^A_{t+1}x^B_t)p(\theta,x^A_{t},Y_t\mid  X^B_{1:t},Y_{1:t-1})d\theta dx^A_{t+1}dx^A_t} \notag\\
    &= \frac{\int_{x^A_t}p(X^B_{t+1}\mid X^B_t,Y_t)p(X^A_{t+1}\mid \Theta,x^A_t,X_t^B) p(Y_{t}\mid \Theta,x^A_t,X^B_{1:t},Y_{1:t-1})p(\Theta,x^A_t\mid X^B_{1:t},Y_{1:t-1})dx^A_t}{\int_{x^A_t}\int_{x^A_{t+1}}\int_{\theta}p(X_{t+1}^B\mid Y_t,X_t^B)p(x^A_{t+1}\mid \theta,x^A_{t+1}x^B_t) p(Y_{t}\mid \theta,x^A_t,X^B_{1:t},Y_{1:t-1})p(\theta,x^A_{t}\mid  X^B_{1:t},Y_{1:t-1}) d\theta dx^A_{t+1}dx^A_t}  \label{eq:belief_proof}
\end{align}
Note that the fourth equation of Eq.\eqref{eq:belief_proof} results from the following relationship 
\begin{align}
    p(X^B_{t+1} \mid X^B_t, Y_t, \Theta, X^A_{t+1}, X^A_t, X^B_{1:t-1}, Y_{t-1}) = p(X^B_{t+1} \mid X^B_t, Y_t),
\end{align}
which holds because $X^B_{t+1}$ is conditional independent of $X^A_{t+1}$, $X^A_t$, $\Theta$, and other historical data, given $X^B_t$ and $Y_t$. 

This allows us to take $p(X^B_{t+1} \mid X^B_t, Y_t)$ out of the integration and further decompose $\beta_{t+1}(\Theta, X^A_{t+1})$ as
\begin{align}
 & \beta_{t+1}(\Theta,X^A_{t+1}) = \frac{\int_{x^A_t}p(X^A_{t+1}\mid \Theta,x^A_t,X_t^B)p(Y_{t}\mid \Theta,x^A_t,X^B_{1:t},Y_{1:t-1})p(\Theta,x^A_t\mid X^B_{1:t},Y_{1:t-1})dx^A_t}{\int_{x^A_t}\int_{x^A_{t+1}}\int_{\theta}p(x^A_{t+1}\mid \theta,x^A_{t+1}x^B_t)p(Y_{t}\mid \theta,x^A_t,X^B_{1:t},Y_{1:t-1})p(\theta,x^A_{t}\mid X^B_{1:t},Y_{1:t-1})d\theta dx^A_{t+1}dx^A_t}\notag\\
    & = \frac{\int_{x^A_t}p(X^A_{t+1}\mid \Theta,x^A_t,X_t^B)\pi^s_t(Y_{t}\mid \Theta,x^A_t,X^B_{1:t},Y_{1:t-1})\beta_t(\Theta,x^A_t)dx^A_t}{\int_{x^A_t}\int_{x^A_{t+1}}\int_{\theta}p(x^A_{t+1}\mid \theta,x^A_{t+1}x^B_t)\pi^s_t(Y_{t}\mid \theta,x^A_t,X^B_{1:t},Y_{1:t-1})\beta_t(\theta,x^A_{t})d\theta dx^A_{t+1}dx^A_t}.
\end{align}

\subsection{Proof of Lemma~\ref{le:bellman equiv}}
\label{appendix:proof for bellman equiv}
According to policy $\pi^s_t$ and belief state $\beta_t$, the Bellman optimality equation \eqref{eq:Bellman optimality equality} can be decomposed as (Since $h_t$ is give, we can simply $\pi^s_t(Y_t\mid \Theta, X^A_t, h_t)$ and $\pi^s(Y_t\mid \Theta, X^A_t)$):
\begin{align}\label{eq:Bellman optimality equality 
decomposed}
V^{*}_t(h_{t}) & = \min_{\pi^{s}_t}\int_{y_t,\theta}\int_{x^A_t}\pi^{s}_t(Y_t\mid\theta,x^A_t)\beta_t(\theta,x^A_t)dx^A_t \log\frac{\int_{x^A_t}\pi^{s}_t(Y_t\mid\theta,x^A_t)\beta_t(\theta,x^A_t)dx_t}{\beta_t(\theta)\int_{\theta}\int_{x^A_t}\pi^{s}_t(Y_t\mid\theta,x^A_t)\beta_t(\theta,x^A_t)dx_td\theta}d\theta dy_t\notag\\ & \quad + \int_{x_t^A,x^A_{t+1},y_t}r(\theta^*,x_t^A,x^A_{t+1},y_t)p(x^A_{t+1}\mid \theta^*,x_t^A) \pi^{s}_t(Y_t\mid\theta^*,x^A_t)\beta_t(\theta^*,x_t^A)dy_tdx^A_{t+1}dx^A_t\notag \\ & \quad +  \lambda \int_{x^A_t,\theta,y_t}\pi^{s}_t(Y_t\mid\theta,x^A_t)\beta_t(\theta,x^A_t)d(x^A_t,y_t)dy_{t} d\theta dx^A_t - \lambda\hat{D}+ \mathbb{E}(V^{*}_{t+1}(h_{t+1}\mid h_{t})),
\end{align}
We define the immediate cost $C_t$ when taking policy set $\pi^s_t$ under history $h_t$ as:
\begin{align}
C_t(h_{t},\pi^t_s) & = \int_{y_t,\theta}\int_{x^A_t}\pi^{s}_t(Y_t\mid\theta,x^A_t)\beta_t(\theta,x^A_t)dx^A_t
\log\frac{\int_{x^A_t}\pi^{s}_t(Y_t\mid\theta,x^A_t)\beta_t(\theta,x^A_t)dx_t}{\beta_t(\theta)\int_{\theta}\int_{x^A_t}\pi^{s}_t(Y_t\mid\theta,x^A_t)\beta_t(\theta,x^A_t)dx_td\theta}d\theta dy_t \notag\\& \quad +\int_{x_t^A,x^A_{t+1},y_t}r(\theta^*,x_t^A,x^A_{t+1},y_t)p(x^A_{t+1}\mid \theta^*,x_t^A)
\pi^{s}_t(Y_t\mid\theta^*,x^A_t)\beta_t(\theta^*,x_t^A)dy_tdx^A_{t+1}dx^A_t \notag\\&
\quad + \lambda \int_{x^A_t,\theta,y_t}\pi^{s}_t(Y_t\mid\theta,x^A_t)\beta_t(\theta,x^A_t)d(x^A_t,y_t) dy_{t} d\theta dx^A_t - \lambda\hat{D}
\end{align}
From each part of $C_t$, we can see that given history $h_t$, $C_t$ is determined by belief state $\beta_t$ and policy $\pi^s_t$, thus we use $C_t(h_{t},\pi^t_s)$ to denote the immediate cost $C_t$ when taking policy set $\pi^s_t$ under history $h_t$.

Based on this dependency, we can use induction to assume that $V^{*}_{t+1}(h_{t+1})$ also depends on belief state $\beta_{t+1}$. Then, we rewrite the Bellman optimality equation of $V^{*}_t(h_{t})$ as:
\begin{align}\label{eq:Bellman optimality equality}
    V^{*}_t(h_{t}) & = \min_{\pi^{s}_t}C_t(h_{t}, \pi^{s}_t)
   + \mathbb{E}(V^{*}_{t+1}(\beta_{t+1}\mid h_{t})).
\end{align}
Using $\beta_{t+1} = \Phi(\beta_t,\pi^s_t,y_t)$ to compactly represent this recursion, we can decompose the second term $\mathbb{E}(V^{*}_{t+1}(\beta_{t+1}\mid h_{t}))$ as:
\begin{align}
    \mathbb{E}(V^{*}_{t+1}(\beta_{t+1}\mid h_{t}))& = \int_{y_t}p(y_t\mid h_{t})V^{*}_{t+1}(\beta_{t+1})dy_t
    \notag\\& = \int_{y_t,\theta,x^A_t}\pi^s_t(y_t\mid\theta,x^A_t)\beta_t(\theta,x^A_t)V^{*}_{t+1}(\Phi(\beta_t,\pi^s_t,y_t))dx^A_td\theta dy_t,
\end{align}
which also depends on $\beta_t$. Thus, combining this term with $C_t(h_{t},\pi^t_s)$ and using induction, we can prove that $ V^{*}_t(h_{t})$ depend on belief state $\beta_t$.

\section{Proofs of Section~\ref{sec: Utility-Aware Privacy-Preserving Data Sharing Framework}}

\label{appendix:weight_update_derivation}

The belief state is then approximated by the weighted sum of these particles:
\begin{align}
    \beta_t(\Theta, X^A_t) &\approx \sum_{i=1}^N \omega_{i,t} \delta\left( (\Theta, X^A_t) - (\theta_{i}, x^A_{i,t}) \right), \label{eq:approx_belief}
\end{align}
where $\delta$ is the Dirac delta function. Eq.\eqref{eq:approx_belief} serves as a discrete weighted approximation to the true posterior $p(\Theta,X^A_t\mid Y_{1:t},X^B_{1:t})$. 

The weights $\{\omega_{i,t}\}_{i=1}^N$ are chosen using the principle of importance sampling. Suppose $p(\Theta,X^A_t\mid Y_{1:t-1},X^B_{1:t})$ is a probability density from which it is difficult to draw samples. We define the easier-to-sample proposal joint distribution (i.e. importance density) of $\theta_{i}$ and $X^A_t$ as  $q(\theta_{i},X^A_{i,t}\mid Y_{1:t-1},X^B_{1:t})$. We get the sample of the new distribution $(\theta_{i},X^A_{i,t})$ and update the weight $\omega_{i,t}$ as follows:

\begin{align}
    \omega_{i,t} \propto \frac{p(\theta_{i},X^A_{i,t}\mid Y_{1:t-1},X^B_{1:t})}{q(\theta_{i},X^A_{i,t}\mid Y_{1:t-1},X^B_{1:t})}
\end{align}

The proposal joint distribution $q(\Theta,X^A_t\mid X^B_{1:t},Y_{1:t-1})$ is chosen to have the a factorized form following the Bayesian rule:
\begin{equation}
\begin{aligned}
    q(\Theta,X^A_t\mid X^B_{1:t},Y_{1:t-1}) = q(X^A_t\mid \Theta,X^A_{t-1},X^B_{1:t},Y_{1:t-1})q(\Theta,X^A_{t-1}\mid X^B_{1:t-1},Y_{1:t-2})
\end{aligned}
\label{eq: q_fac}
\end{equation}

Similarly, the joint density $p(\Theta,X^A_{t}\mid Y_{1:t},X^B_{1:t})$ can be factorized as:
\begin{align}
\label{eq: p_fac}
    p(\Theta,X^A_{t}\mid Y_{1:t},X^B_{1:t}) &= p(\Theta,X^A_{t-1}\mid X^B_{1:t-1},Y_{1:t-2})p(X^B_{t}\mid X^B_{t-1},Y_{t-1})p(X^A_t\mid \Theta,X^A_{t-1},X^B_{t-1})\notag\\
    &\quad p(Y_{t-1}\mid \Theta,X^A_{t-1},X^B_{1:t-1},Y_{1:t-2})\big/p(X^B_{t},Y_{t-1}\mid X^B_{1:t-1},Y_{1:t-2})\notag\\
    & \propto p(\Theta,X^A_{t-1}\mid X^B_{1:t-1},Y_{1:t-2})p(X^B_{t}\mid X^B_{t-1},Y_{t-1})p(X^A_t\mid \Theta,X^A_{t-1},X^B_{t-1})\notag\\ & \quad p(Y_{t-1}\mid \Theta,X^A_{t-1},X^B_{1:t-1},Y_{1:t-2})
\end{align}

Using the factorized form in Eqs.~\eqref{eq: q_fac} and \eqref{eq: p_fac}, we can derive the recursive weights update scheme as follows:
\begin{align}
\label{eq:weight update process}
    \omega_{i,t} \propto \frac{\omega_{i,t-1}p(X^B_{t}\mid X^B_{t-1},Y_{t-1})p(Y_{t-1}\mid \theta_{i},X^A_{i,t-1},X^B_{1:t-1},Y_{1:t-2})p(X^A_{i,t}\mid \theta_{i},X^A_{i,t-1},X^B_{t-1})}{q(X^A_{i,t}\mid \theta_i,X^A_{i,t-1},X^B_{1:t},Y_{1:t-1})}
\end{align}

Assume that the important density $q(X^A_{i,t}\mid \theta_i,X^A_{i,t-1},X^B_{1:t},Y_{1:t-1}) = p(X^A_{i,t}\mid \theta_i,X^A_{i,t-1}, X^B_{1:t-1})$, the final process of weights update can be concluded
\begin{align}
    \omega_{i,t} \propto \omega_{i,t-1}p(X^B_{t}\mid X^B_{t-1},Y_{t-1})p(Y_{t-1}\mid \theta_{i},X^A_{i,t-1},X^B_{1:t-1},Y_{1:t-2}).
\end{align}

\section{Experiment Details}
\label{appendix: Experiment}
The parameters of the scenario settings and training settings are included in Table~\ref{tb:notations}

\begin{table}[ht]
\centering
\caption{Value of Parameters}
\label{tb:notations}
\begin{tabular}{lll}
\toprule
\textbf{Category} & \textbf{Parameter and Definition} & \textbf{Value} \\ \midrule
\multicolumn{3}{l}{\textbf{Simulation Parameters}} \\ \midrule
$\Delta t$ & Time interval to share information & $0.2\,s$ \\
$(s^{*},v^{*})$ & Equilibrium spacing and velocity & $(20\,\mathrm{m}, 15\,\mathrm{m/s})$ \\
$(s_{st}, s_{go})$ & Thresholds of spacing & $(5\,\mathrm{m}, 35\,\mathrm{m})$ \\
$(a_{min}, a_{max})$ & Thresholds of acceleration & $(-5\,\mathrm{m/s}^2, 5\,\mathrm{m/s}^2)$ \\
$(\mu_1, \eta_1)$ & AV's feedback gains of linear controller & $(0.1, 0)$ \\
$(\mu_2, \eta_2)$ & CAV's feedback gains of linear controller & $(-0.5, 0.5)$ \\
$(\mu_3, \eta_3)$ & HDV's feedback gains of linear controller & $(-0.2, 0.2)$ \\
$v_{\text{max}}$ & Maximum velocity & $30\,\mathrm{m/s}$ \\
$w(t)$ & Mean of the velocity fluctuation of the AV & $1\,\mathrm{m/s}^2$ \\

\midrule
\multicolumn{3}{l}{\textbf{Adversaries Parameters}} \\ \midrule
$(\sigma_v, \sigma_s)$ & Std. deviations of the observation noise & $(0.3, 0.3)$ \\
$\lambda$ & Forgetting factor & $0.99$ \\
$\delta$ & Initial confidence parameter & $1$ \\ \midrule
\multicolumn{3}{l}{\textbf{Training Parameters}} \\ \midrule
$L$ & Time horizon of each episode & $200$ \\
$\hat{D}$ & Upper bound for expected total deviation & $800$ \\
$\lambda$ & Lagrangian multiplier & $1$ \\
$\rho$ & Weight of privacy measure & $1$ \\
$D^v$ & Number of dimensions of the learned location vector  & $10$ \\
$lr_a$ & Actor learning rate & $1e\text{-}3$ \\
$lr_c$ & Critic learning rate & $1e\text{-}3$ \\
Actor layers & Number of layers in the actor network & $2$ \\
Critic layers & Number of layers in the critic network & $2$ \\
MGF layers & Number of layers in the MGF network & $1$ \\
\bottomrule
\end{tabular}
\end{table}

\subsection{Setup Details}
\subsubsection{Scenarios Setting}
\label{appendix: simulation setting}
\noindent\textbf{Modeling of Mixed-Autonomy Platoon}: For the leading vehicle, we utilize a random acceleration disturbance setup, with a Gaussian distribution with a mean of $0$ and a variance of $1$.

For the mixed-autonomy platoon containing the HDV and the CAV, the system dynamics can be written as:
\begin{subequations}
\label{eq:system_dynamics}
\begin{align}
    \dot{s}_{i}(t) &= 
    \begin{cases} 
        v_{i-1}(t) - v_{i}(t), & \quad i = 1, 2, \\ 
        v_{i-1}(t) - v_{i}(t) + g_s(t), & \quad i = 3,
    \end{cases} \\
    \dot{v}_{i}(t) &= 
    \begin{cases} 
        u(t), & \quad i = 2, \\ 
        \mathbf{F}_{\Theta}\left(s_i(t), v_i(t), v_{i-1}(t)\right) + g_a(t), & \quad i = 3.
    \end{cases}
\end{align}
\end{subequations}
where the acceleration rate of CAV is determined by the control action $u(t)$, and the acceleration rate of HDV is governed by the car-following model $a(t) =\mathbf{F} _{\Theta}$, which depends on its current speed $v_3(t)$, the preceding vehicle's speed $v_2(t)$, and the spacing $s_3(t)$. To simulate the variability observed in real-world human driving behavior, we introduce additional disturbances into both the car-following model and the spacing dynamics of HDV. Specifically, the acceleration update includes a disturbance $g_a(t) \sim \mathcal{N}(0,0.1)$, while the spacing update incorporates a disturbance $g_s(t) \sim \mathcal{N}(0,0.5)$.

In detail, for the CAV, upon receiving information from all vehicles, the CAV control action $u(t)$ is determined using a distributed linear controller following \cite{wang2021leading}, aiming to maintain an equilibrium state $\left(s^{\star},v^{\star}\right)$:
\begin{align}
    u(t) = \sum_{i=1}^3\left(\mu_{i}\left(s_{i}-s^{\star}\right) + \eta_{i}\left(v_{i}-v^{\star}\right)\right)
\end{align}
where $\mu_{i}$ and $\eta_{i}$ are feedback gains based on string stability conditions.

The following HDV’s car-following behavior $\mathbf{F}_{\Theta}(\cdot)$ is parameterized by sensitive parameter $ \Theta$. Using the Full Velocity Difference Model (FVD)~\cite{jiang2001full} as an example, which can be represented as:
\begin{align}\label{eq:FVD}
     \mathbf{F} (\cdot) = m\left(V_{\text{FVD}}\left(s_3(t)\right)-v_3(t)\right) + n (v_{2}(t)-v_{3}(t))
\end{align}
where parameters $m$ and $n$ represent sensitivity to spacing and velocity errors, respectively. In this study, the sensitive parameters $\Theta$ (i.e., $m$ and $n$) are fixed throughout the simulation. $V_{\text{FVD}}(s)$ is the spacing-dependent desired velocity, denoted as:
\begin{equation}
V_{\text{FVD}}(s)= 
\begin{cases}
0, & s \leq s_{\mathrm{st}} \\
\frac{v_{\max }}{2}\left(1-\cos \left(\pi \frac{s-s_{\mathrm{st}}}{s_{\mathrm{go}}-s_{\mathrm{st}}}\right)\right), & s_{\mathrm{st}} < s < s_{\mathrm{go}} \\ 
v_{\mathrm{max}}, & s \geq s_{\mathrm{go}}
\end{cases}
\end{equation}
where $s_{\mathrm{st}}$ and $s_{\mathrm{go}}$ are spacing thresholds, and $v_{\mathrm{max}}$ is the maximum velocity.

\noindent\textbf{Privacy-Preserving Framework Setting}: 

\emph{1. Data Distortion Measurement}: Data distortion measurement is defined as the Euclidean distance between the distorted and true values of speed and spacing:
    \begin{align}
    d(X^A_t,Y_t) = \sqrt{\left(v_t-\Tilde{v}_t \right)^2 +\left(s_t-\Tilde{s}_t\right)^2}
\end{align}
where $X^A_t = (v_t,s_t)$ are true state data of data provider, and $Y_t = (\tilde{v}_t,\tilde{s}_t)$ are distorted state data.

 \emph{2. Control Performance}: The fuel consumption rate $f_i$(mL/s) for the $i$-th vehicle is calculated using the instantaneous model proposed in \cite{bowyer1985guide}:
\begin{equation}
f_i = \begin{cases} 
    0.444 + 0.090 R_i v_i + \left[0.054 a_i^2 v_i\right]_{a_i>0}, & R_i > 0 \\ 
    0.444, & R_i \leq 0 
\end{cases}
\end{equation}
where $R_i = 0.333 + 0.00108 v_i^2 + 1.200 a_i$, with $v_i$ as the vehicle’s velocity and $a_i$ as its acceleration.

\noindent\textbf{Adversaries Setting}: 

\emph{1. Bayesian Inference}: The adversary continuously updates its estimate of the $\Theta$'s distribution (i.e., belief state $\beta_t$) based on prior knowledge and shared data using PF. When inferring the system parameters using distorted data, to reflect a worst-case estimate of data leakage, we still assume all conditional probabilities represented by $p(\cdot\mid\cdot)$ (i.e., state transition, data distortion policy) are public knowledge to both the data provider and data user. 

When inferring the system parameter using real data, we assume that the Bayesian attacker employs an observer capable of updating the belief about $\Theta$. In other words, the received data is treated as noisy measurements of the HDV's state. The observer calculates the likelihood of each particle and updates their weights accordingly. Let the state data observed at time $t$ be $O_t = (o_{t,1}, o_{t,2})$, the discrepancy between the particles state and observed states is measured as $\text{errors}_i = [s_{i,t} - o_{t,1}, v_{i,t} -o_{t,2}]$,
where $s_{i,t}$ and $v_{i,t}$ denote the $i$-th particle’s velocity and position. The likelihood for each particle’s predicted state, assuming Gaussian observation noise, is:
\begin{align}
    p(O_t | x_{i,t}) = \exp\left(-2 \left( \frac{\text{errors}_{i,1}^2}{\sigma_{v}^2} + \frac{\text{errors}_{i,2}^2}{\sigma_{s}^2} \right)\right),
\end{align}
where $\sigma_{v} = 0.3$ and $\sigma_{s} = 0.3$ are the standard deviations of the observation noise for velocity and position, respectively. The factor $-2$ adjusts the sensitivity of the likelihood to the error magnitude. The Bayesian attacker then updates the weights of each particle:
\begin{align}
    w_i^{\text{new}} = \frac{p(O_t | x_{i,t}) \cdot w_i^{\text{old}}}{\sum_{j} p(O_t | x_{j,t}) \cdot w_j^{\text{old}}},
\end{align}
allowing the Bayesian attacker to iteratively refine its estimate based on real data.

\noindent \emph{2. Recursive Least Squares}: The RLS filter iteratively estimates $\Theta$ by adjusting model parameters based on shared state data. The filter is initialized with a parameter vector $\theta_{\text{RLS}}$ and an inverse covariance matrix $\mathbf{P}$ set to $\delta^{-1} \mathbf{I}$, with $\delta$ for initial confidence. Each iteration updates $\theta_{\text{RLS}}$ using the gain vector $\mathbf{K}$ and forgetting factor $\lambda$:
\begin{align}
    \mathbf{K} = \frac{\mathbf{P} x}{\lambda + x^T \mathbf{P} x}, \quad \theta_{\text{RLS}} = \theta_{\text{RLS}} + \mathbf{K} \left( y - x^T \theta_{\text{RLS}} \right)
\end{align}
where $y$ is the new observation. $\mathbf{P}$ is updated as:
\begin{align}
    \mathbf{P} = \frac{\mathbf{P} - \mathbf{K} x^T \mathbf{P}}{\lambda}
\end{align}

In our research, the FVD model in~Eq.\eqref{eq:FVD} can be approximated as a linear system, where $x=m(V_{\text{FVD}}(s_3(t))-v_3(t)), n(v_{2}(t)-v_{3}(t))) $ and $y=\dot{v}_{3}(t)$. This formulation enables an adversary to iteratively refine its estimates of $m$ and $n$, thereby progressively inferring the HDV's driving characteristics from the shared data.

\subsubsection{Solution Method Setting}
\label{appendix: paricle formulation setting}

\noindent\textbf{Sequential Importance Resampling}: Degeneracy refers to the phenomenon where, after several iterations, only a few particles have significant weights, while the majority carry negligible weights, reducing the diversity of particles and leading to poor estimation accuracy. The effective sample size $\hat{N}_{\text{eff}}$ determines when to resample, which can be estimated from the normalized particle weights $\tilde{\omega}_{i,t}$:
\begin{align}
    \hat{N}_{\text{eff}} = \frac{1}{\sum_{i=1}^N (\tilde{\omega}_{i,t})^2}
\end{align}

We set a constant threshold $108$, which represents $ \frac{1}{3}$ of particles. If $\hat{N}_{\text{eff}} \leq 108$, indicating significant degeneracy, a resampling is performed to maintain the diversity of the particles. The resampling steps include two steps. First, A new set of particles $\{X_{i,t}\}_{i=1}^N$ is drawn with replacement from the current set, with each particle selected based on the normalized weights $\tilde{\omega}_{i,t}$. Second, after resampling, all particles are assigned equal weights: $\omega_{i,t} = \frac{1}{N}, \quad \text{for } i = 1, 2, \dots, N$. This process prevents the dominance of a single particle and ensures that the Particle Filter remains effective over time.

\noindent \textbf{MGF Encoder}: The Moment Generating Function (MGF) is defined as the expectation of the random variable. For particle beliefs, the MGF of $\beta_t$ is represented as:
\begin{align}
    M_{\beta_t}(v) = \sum_{i=1}^N \omega_{i,t} e^{v^\top h_{i,t}},
\end{align}
where $h_{i,t} = (\theta_{i,t}, x_{i,t})$, and $v \in \mathbb{R}^{D_v}$ represents a vector of locations learned by the MGF encoder.

To encode the particle belief $\{(\theta_{i,t}, x_{i,t}), \omega_{i,t}\}_{i=1}^N$, we extract the features $(\bar{h}_t, M_{\beta_t}^{1:m})$. Here, $\bar{h}_t = \sum_{i=1}^N \omega_{i,t} h_{i,t}$ is the weighted mean of the particles, representing the first-order moment of the particle belief. $M_{\beta_t}^{1:m}$ is a vector $(M_{\beta_t}^1, M_{\beta_t}^2, \ldots, M_{\beta_t}^m)$, where each $M_{\beta_t}^i$ is computed as $M_{\beta_t}(v^i)$, representing the $i$-th MGF feature at the learned location $v^i$. These features capture higher-order moments of the particle belief, efficiently encoding the belief into a low-dimensional vector.

We use a fully connected layer with the number of output features equal to the MGF features. The activation function applied is ReLU. The processed MGF features are weighted using an exponential function applied to the particle weights.

\noindent\textbf{Actor Network and Critic Network}:  The network includes a 1-layer MLP for the MGF encoder and 2-layer MLPs for the actor and critic networks. The output of the policy, i.e., the distorted data $Y_t$, takes value from a discretized set of $121$ states (i.e., $11$ values for speed and spacing, respectively). Such discretization can accelerate the convergence of training and further protect privacy of HDVs. The actor and critic networks take the belief summary (the output of the MGF encoder) and additional state information as input. 

Specifically, for the actor network, the merged data $(\bar{h}_t, M_{\beta_t}^{1:m})$ is concatenated with the state input $(\theta,x^A_t)$ and passed through two fully connected layers. The first layer has 64 hidden units and applies ReLU activation, followed by another layer that outputs the parameters of a Dirichlet distribution using the Softplus function to ensure non-negative values. The output dimension is set to match the discrete distorted data set which is 121. The reason we require the actor network to output the parameters of a Dirichlet distribution is that these parameters effectively control the shape of the data-sharing policy, providing greater generality and flexibility.

For the critic network, the merged data $(\bar{h}_t, M_{\beta_t}^{1:m})$ are directly processed through fully connected layers with ReLU activation. The first layer also has 64 hidden units and the output layer produces a single scalar value representing the state-value function.

\end{document}